\documentclass[aps,pre,twocolumn,showpacs,floatfix,amsmath,
  superscriptaddress]{revtex4}
\usepackage{epsfig}

\begin{document}

\title{
Wigner--Dyson statistics for a class of integrable models
}

\author{L. Benet}
\affiliation{Centro de Ciencias F\'{\i}sicas, U.N.A.M., 
  Apdo. Postal 48--3, 62251 Cuernavaca, Mor., M\'exico}
\affiliation{Centro Internacional de Ciencias A.C., 
  Apdo. Postal 6-101, 62131 Cuernavaca, Mor., M\'exico}
\affiliation{LPTMS, Universit\'e Paris--Sud, Centre Scientific d'Orsay,
  B\^at. 100, F--91405 Orsay Cedex, France.}
\author{F. Leyvraz}
\affiliation{Centro de Ciencias F\'{\i}sicas, U.N.A.M., 
  Apdo. Postal 48--3, 62251 Cuernavaca, Mor., M\'exico}
\author{T.H.Seligman}
\affiliation{Centro de Ciencias F\'{\i}sicas, U.N.A.M., 
  Apdo. Postal 48--3, 62251 Cuernavaca, Mor., M\'exico}
\affiliation{Centro Internacional de Ciencias A.C., 
  Apdo. Postal 6-101, 62131 Cuernavaca, Mor., M\'exico}

\date{\today}

\begin{abstract}
We construct an ensemble of second--quantized Hamiltonians with two
bosonic degrees of freedom, whose members display with probability one
GOE or GUE statistics. Nevertheless, these Hamiltonians have a second
integral of motion, namely the boson number, and thus are
integrable. To construct this ensemble we use some ``reverse
engineering'' starting from the fact that $n$--bosons in a two--level
system with random interactions have an integrable classical limit by
the old Heisenberg association of boson operators to actions and
angles. By choosing an $n$--body random interaction and degenerate
levels we end up with GOE or GUE Hamiltonians. Ergodicity of these
ensembles completes the example.
\end{abstract}


\maketitle

Recently, there has been considerable interest in spinless $n$--boson
systems with $k$--body random interactions~\cite{Kota01,BW03}, both for
degenerate~\cite{Asaga01} and non--degenerate~\cite{Patel00,Chavda03}
single--particle levels. In particular, anomalous statistics for the
two--level system have been understood from the fact that these
systems are integrable~\cite{BJL03}. The two--level ensemble
corresponds to a time--independent two degrees of freedom system, in
the sense that creation and annihilation operators for the two
single--particle levels are canonical operators for the system. The
second integral of motion corresponds to the conservation of the
number of bosons. The Hamiltonian is a function of the creation and
annihilation operators, whose precise form depends on the question
whether we have two--body interactions or more complicated many--body
interactions. The number of (random) coefficients in this model
increases quadratically with the rank $k$ of the interaction. A
classical Hamiltonian can formally be written for any number of
particles but in fact only for large particle number the quantum
problem reaches the classical limit~\cite{heis25,Yaffe82}. Thus the
boson number plays the role of an action. For fixed particle number,
the model can be reduced to a Hamiltonian of one degree of freedom 
with the number of bosons appearing as parameter.

In this paper we shall consider a different large--$n$ limit for the
two--level system. Our model consists in choosing the rank of the
interaction equal to the particle number, $k=n$. Hence the Hamiltonian
changes for different values of the particle number. This choice, by
definition of the ensemble, directly leads to a GOE or GUE
matrix~\cite{mehta}. We then consider the limit of large particle
number. Clearly, we will have an ensemble of systems which have GOE or
GUE statistics with probability one~\cite{pandey}, while being
integrable in a well--defined limit of large actions. This represents
an important caveat concerning the idea that Wigner--Dyson statistics
are characteristic of classical chaotic systems. Note that this has no
bearing on the fairly--well established quantum chaos
conjecture~\cite{BGS84,MacDonald,casati,Berry81,Berry85,Andr95,Leyvraz},
which establishes that classical chaos implies typically Wigner--Dyson
statistics. In a similar vein, Wu {\em et al}.~\cite{wu} have
performed the following calculation: after having chosen an unfolded
{\em fixed\/} spectrum of given length $N$ generated via the
diagonalization of a matrix taken from the GOE, they fit a
one--dimensional potential, the spectrum of which coincides with the
above spectrum for its first $N$ values. The main difference between
their work and ours is, that we display explicitly an ensemble of
integrable Hamiltonians having a GOE spectrum.

The Hamiltonian in second quantization is
\begin{equation}
\label{eq1}
{\hat H} = \sum_{s,t=0}^k \frac{v_{s,t}}{{\cal N}_{s,t}}
 (b_1^\dagger)^{s} (b_2^\dagger)^{k-s} (b_1)^{t} (b_2)^{k-t} \ .
\end{equation}
Here, $b_j^\dagger$ creates a spinless boson in single--particle
level $j$ $(j=1,2)$, $b_j$ destroys it, and the $k$--body matrix
elements $v_{s,t}$ corresponds to a GOE or a
GUE~\cite{Asaga01}. The number operator ${\hat n}=b_1^\dagger
b_1+b_2^\dagger b_2$ commutes with the Hamiltonian independently of
the rank $k$ of the interaction. The combinatorial factors ${\cal
N}_{s,t}=[s! (k-s)! t! (k-t)!]^{1/2}$ in Eq.~(\ref{eq1})
are introduced in order to have {\it exactly} a GOE or GUE Hamiltonian
for $n=k$, where $n$ is the number of bosons~\cite{Kota01,BW03}. For
the two--level system, the dimension of Hilbert space is  
$N=n+1$.

The classical Hamiltonian is obtained as 
follows~\cite{jacob99,BJL03}: ${\hat H}$ is symmetrized with 
respect to the ordering of the creation and annihilation operators. This 
permits a correct assignment of the zero--point energy and therefore a 
 one--to--one classical--quantum energy comparison. Then, 
we use Heisenberg semiclassical rules~\cite{heis25}
\begin{equation}
\label{eq2}
b_j^\dagger \longrightarrow I^{1/2}_j \exp( i \phi_j), \quad
b_j         \longrightarrow I^{1/2}_j \exp(-i \phi_j) .
\end{equation}

Formally, the classical Hamiltonian can be written as $H=H_0+V$. The
unperturbed Hamiltonian $H_0$, which includes a constant associated
with the zero--point energy correction, depends only upon the actions
$I_1$ and $I_2$; the residual interaction $V$ depends also on the
angles $\phi_1$ and $\phi_2$. These terms are explicitly given by
\begin{eqnarray}
\label{eq3}
H_0 & = & \sum_{s=0}^k \frac{v_{s,s}}{{\cal N}_{s,s}} \, 
    P_s[I_1-\frac{1}{2},s] \, P_{k-s}[I_2-\frac{1}{2},k-s]  , \\
\label{eq4}
V & = & \sum_{s>t} \frac{v_{s,t} (I_1 I_2)^{(s-t)/2}}{{2 \, \cal N}_{s,t}} 
    \cos[(s-t)(\phi_1-\phi_2)] \nonumber \\
  & \times & \bigl( P_t[I_1-\frac{1}{2},s] + P_t[I_1-\frac{1}{2},t] \bigr) 
      \nonumber \\
  & \times & \bigl( P_{k-s}[I_2-\frac{1}{2},k-s] + 
       P_{k-s}[I_2-\frac{1}{2},k-t] \bigr) .
\end{eqnarray}
In these equations (details of their derivation will be given
elsewhere), $P_t[I,s]$ is a polynomial of degree $t$ on $I$ defined as
($s\ge t\ge 0$)
\begin{equation}
\label{eq5}
P_t [I,s] = \prod_{i=1}^{t} [I - (s-i)] \ .
\end{equation}
Here we have assumed real matrix elements $v_{s,t}$, which are
independent random variables, Gaussian distributed with zero 
mean and variance given by $\overline{v_{s,t} v_{s',t'}}=\delta_{s s'}
\delta_{t t'} + \delta_{s t'} \delta_{t s'}$, with the overline
denoting ensemble average.
Hence the $k$--body interaction matrices form a GOE. The analogous
case for the GUE is obtained by replacing in Eq.~(\ref{eq4}) $v_{s,t}$
by $|v_{s,t}|$ and introducing a phase $\omega_{s,t}$ in the cosine
function.

 The Hamiltonian is
written as a polynomial of the two actions with random coefficients,
and cosines of the difference of the two angles. The
invariant is given by $K = I_1 + I_2 = n+1$ which reflects the
translational symmetry of the interaction with respect to the
angles. Again we can test the corresponding Poisson
brackets~\cite{BJL03}. This form displays explicitly that we
deal with an integrable two--degrees of freedom problem. If we fix the
invariant, i.e. the particle number $n$, we can reduce
the Hamiltonian to one degree of freedom. The quantum
Hamiltonian acts on a Hilbert space of dimension
$n+1$, which is the number of ways we can distribute the $n$ bosons in
the two single--particle levels. Formally, this corresponds to the
so--called ``polyads'' for algebraic Hamiltonians in molecular
systems~\cite{jacob99}. The reduced Hamiltonian is obtained by
performing the canonical transformation defined by the generating
function $W = K \phi_1 + J (\phi_2-\phi_1)$. The new actions $K$ and
$J$ and their corresponding canonically conjugated angles $\chi$ and
$\psi$ are related to the old variables by
\begin{subequations}
\label{eq6}
\begin{eqnarray}
\label{eq6a}
I_1 & = K - J , \qquad I_2 & = J , \\
\label{eq6b}
\chi & = \phi_1 , \qquad \qquad \psi & = \phi_2-\phi_1 .
\end{eqnarray}
\end{subequations}
Substituting the new variables in Eqs.~(\ref{eq3})
and~(\ref{eq4}), and fixing $K$  yields explicitly the
reduced Hamiltonian.

We now have established the main point of the paper: if we
consider $k=n$, by construction
the quantum Hamiltonian coincides with a
GOE~\cite{Kota01,Asaga01}. Each member of the ensemble is associated,
in the limit of large $n(=k)$, to a classical Hamiltonian with one
effective degree of freedom. This defines the ensemble of integrable
Hamiltonians.

\begin{figure}
\includegraphics[angle=90,width=8.5cm]{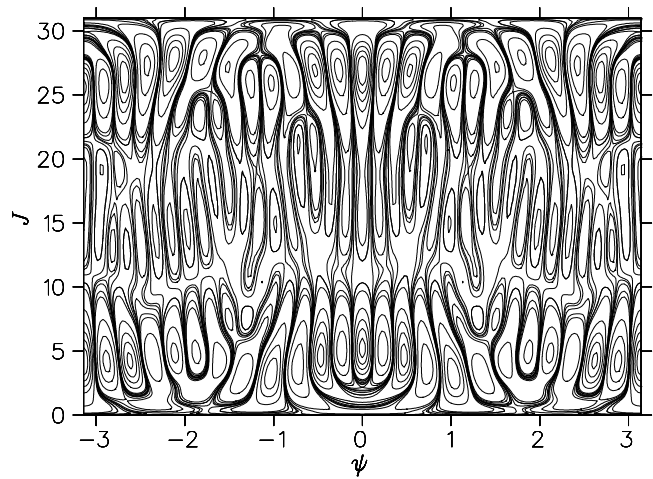}
\caption{\label{fig1} Phase--space structure of a specific realization
of the reduced Hamiltonian for $k=n=30$. }
\end{figure}

Let us now qualitatively understand why
this happens. For this purpose we investigate the structure of the
phase space of a typical member of the ensemble for fixed $n\gg
1$. A Poincar\'e surface of the two degrees of
freedom system constructed for constant $K=n+1$ is equivalent to a
contour plot of the reduced one--dimensional Hamiltonian. 
Fixing $K$ makes for a compact phase space in the
reduced coordinates $\psi$ and $J$ defined above. The volume
corresponds to $n+1$ as we normalized to one quantum state
per unit cell. 
The 
number of random coefficients entering in the Hamiltonian grows 
quadratically on $k$. 
The equations of
motion then contain polynomials in $J$ and $K-J$ (of degree up
to $k=n$) multiplied by sines or cosines of multiples of
$\psi$.  In this case, the degree of the polynomials 
sets $\sim n^2$ as the upper limit for the total number of fixed points 
(stable and unstable). While a more precise estimate for their number  
is difficult to establish, our numerical results 
show that the fixed points proliferate more rapidly 
than $n$, and come close to the upper bound. 
The typical phase--space portrait will 
therefore consist of elliptic islands surrounded by separatrices in a 
complicated mesh. In Fig.~\ref{fig1} we illustrate the phase--space 
structure of a member of the ensemble for $k=n=30$. The periodic orbits 
whose action satisfy the EBK quantization condition
\begin{equation}
\label{eq7}
S(E_i) \equiv \frac{1}{2\pi} \oint J(E_i) {\rm d}\psi 
  = \frac{1}{2\pi} (n_i + \frac{\alpha_i}{4}),
\end{equation}
define implicitly the $i$--th energy level $E_i$. Here, $n_i$ is an
integer and $\alpha_i$ is the Maslov index of the orbit. Note that
there is {\it a priori} no monotonic dependence of the value of the
action as a function of energy for the reduced Hamiltonian. That is,
for a given energy two or more distinct invariant tori may exist in
different regions of phase space; each of these tori may have a
different value of the corresponding action.

As mentioned above, the number of stable and unstable fixed points
proliferates very rapidly for growing $k=n$. This is a consequence of
the many--body character of the interaction $V$ in Eq.~(\ref{eq4}). In
particular, in the limit we consider, this number grows much faster
than $n\sim \hbar^{-1}$; this implies that the phase--space volume
surrounding each stable fixed point shrinks for $n=k\to\infty$. 

For one degree of freedom systems in the large $n$ limit
with $k$ fixed the spectrum is constructed from sequences of levels
obtained by torus quantization around the elliptic fixed points, {\em i.e.}
most periodic orbits which satisfy the EBK condition can be uniquely 
assigned to one elliptic fixed point.  The
spectrum is thus a superposition of picket--fence 
(equidistant) sequences of levels as reported in~\cite{Asaga01} for 
small values of $k$. For fixed (sufficiently large) values of $k$
we expect a Poisson spectrum to arise on 
short length scales. This expectation is based on the fact that WKB
theory yields a superposition of many picket fence (equidistant) spectra
with different spacings. 

In contrast, in the present case ($k=n$) we have $\sim n^2$ elliptic 
fixed points and only $n+1$ energy levels. Therefore, a stable periodic 
orbit that fulfills the EBK condition~(\ref{eq7}), instead of surrounding 
an elliptic point, has to accommodate and explore more extended regions 
in phase space, coming close to the separatrix associated with many 
different unstable fixed points. Stated in a different way, while the 
periodic orbits of the Hamiltonian are strictly stable, wave packets 
started on initial conditions separated by a distance of the order 
$\hbar^{1/2} \sim n^{-1/2}$ will sample very different regions of 
phase space. This coarse--graining will effectively mimic unstable 
motion. This fact makes plausible that we do not obtain a smooth 
spectrum that can be unfolded to yield (superpositions of) picket--fence 
spectra. Indeed, we find Random Matrix spectral 
fluctuations. Note that the results on the non--ergodic behavior in 
the dense limit ($k$ fixed) of the bosonic $k$--body embedded 
ensembles~\cite{Asaga01} cease to apply in this case by the same 
line of reasoning, and therefore we recover the standard ergodicity of 
the GOE~\cite{pandey}.

Finally, we consider the extension to the GUE case. We recall that in
action and angle variables of the harmonic oscillator, time--reversal
invariance is tested by the invariance of the Hamiltonian under the
transformation $\phi_j\to -\phi_j$ (for $j=1,2$). As mentioned above,
when considering complex $k$--body matrix elements, Eq.~(\ref{eq4})
requires certain modifications; namely, $v_{s,t}$ is replaced by its
modulus $|v_{s,t}|$ and the phases $\omega_{s,t}$ are introduced in
the argument of the cosine functions. The introduction of the
phases $\omega_{s,t}$ in the cosine functions of Eq.~(\ref{eq4})
breaks the time--reversal invariance.

Summarizing, we have defined an ensemble of integrable Hamiltonians of
two degrees of freedom, or equivalently, an ensemble of one degree of
freedom Hamiltonians by fixing the number of bosons $n$. In the limit
$n=k\to\infty$, each member of the (reduced) ensemble is mapped exactly 
by the quantization onto a member of the GOE or GUE. Therefore, the
fluctuation properties of its eigenvalues follow the predictions of
Random Matrix Theory. This result can be interpreted in two ways:
First, the limiting Hamiltonian is considered to be a classical one,
or second, such a limit is not accepted as a classical one. In either
case, for this family of Hamiltonians Wigner--Dyson fluctuations does
not imply chaos in classical dynamics.

We acknowledge financial support by the projects DGAPA (UNAM)
IN--109000, IN--112200 and CONACyT 32173-E. LB was supported by the
PSPA--DGAPA (UNAM) program and thanks O. Bohigas for discussions and
the kind hospitality at the LPTMS (Orsay).

\end{document}